\documentclass[a4paper,10pt]{IEEEtran}

\usepackage{xcolor}
\usepackage{balance,flushend}
\def\BibTeX{{\rm B\kern-.05em{\sc i\kern-.025em b}\kern-.08em
    T\kern-.1667em\lower.7ex\hbox{E}\kern-.125emX}}
\ifCLASSINFOpdf
  \usepackage[pdftex]{graphicx}
\else
  \usepackage[dvips]{graphicx}
\fi

\ifCLASSOPTIONcompsoc
 \usepackage[caption=false,font=normalsize,labelfont=sf,textfont=sf]{subfig}
\else
 \usepackage[caption=false,font=footnotesize]{subfig}
\fi

\begin{document}
\title{Dual Dielectric Metasurface for Simultaneous Sensing and Reconfigurable Reflections}

\author{\IEEEauthorblockN{Mahesh Birari$^1$, Deepak Singh Nagarkoti$^2$, Anestis Katsounaros$^{3,1}$, Hruday Kumar 
Reddy Mudireddy$^2$, \\Jagannath Malik$^4$, and George C. Alexandropoulos$^{5,6}$ 
} \\    
\IEEEauthorblockA{$^1$EpsilonR, UK, $^2$Texas Instruments, USA, $^3$Space Innovation Solutions, Greece\\ $^4$Indian Institute of Technology, Patna, India \\$^5$Department of Informatics and Telecommunications, National and Kapodistrian University of Athens, Greece \\$^6$Department of Electrical and Computer Engineering, University of Illinois Chicago, USA}\\
\IEEEauthorblockA{emails: birarimahesh@gmail.com, nagarkoti@ieee.org, anestis@spaceis.eu, malik@iitp.ac.in, alexandg@di.uoa.gr}
}

\maketitle
\begin{abstract}
This paper presents a novel dual-functional hybrid Reconfigurable Intelligent Surface (RIS) for simultaneous sensing and reconfigurable reflections. We design a novel hybrid unit cell featuring dual elements, which share the same phase center, to support both intended functionalities, with the antenna being miniaturized via a high dielectric material approach. The hybrid unit cell has a size of one eighth of the wavelength forming the foundation of an innovative metasurface that incorporates a sub-wavelength reflecting array of split-ring unit cells integrated with a load-tuning matrix. In particular, two interleaved sensing arrays of half-wavelength spacing, orthogonal polarization, and quarter-wavelength offset are embedded within the proposed dual-functional RIS, each tasked to sense the channel parameters towards one of the end communication nodes wishing to profit from the surface's reconfigurable reflections. Our full-wave simulations, indicatively centered around the frequency of $5.5$ GHz, showcase the promising performance of both designed hybrid unit cells and reflective split-ring ones.
\end{abstract}


\section{Introduction}
Reconfigurable Intelligent Surfaces (RISs) constitute one of the emerging physical-layer technologies for the upcoming sixth Generation (6G) of wireless networks, offering over-the-air wave propagation control~\cite{HZA_TWC_2019,CPH_TWC_2022,RISsurvey2023}, which is lately being leveraged for various applications, including Integrated Sensing And Communications (ISAC)~\cite{RIS_ISAC}, secrecy~\cite{PLS_Kostas}, simultaneous localization and radio mapping~\cite{kim2023ris}, seamless localization and sensing in smart cities~\cite{CKA_COMMAG_2023}, and digital twinning~\cite{MHA_IoTMAG_2023}. However, the optimization of RIS operations comes with certain challenges (e.g., channel estimation and reflective beamforming design~\cite{Tsinghua_RIS_Tutorial}), with most of them resulting from its initially envisioned hardware architecture including almost passive unit cells and a basic controller interfacing the metasurface with the interested end communication nodes~\cite{RIS_challenges}.

The authors in~\cite{alexandropoulos2020hardware} were the first to propose a metasurface hardware architecture incorporating reception Radio Frequency (RF) chains, which can be fed with the impinging signals on the RIS unit cells, when these cells are tuned in a fully absorption state. This functionality was conceptually realized by connecting each cell or group of cells with sampling waveguides to forward the impinging signal(s) to a baseband unit embedded in the RIS controller for further processing (e.g., explicit or implicit channel estimation). In an alternative direction, the authors in~\cite{taha2021enabling} proposed to replace some of the tunably reflecting unit cells of conventional RISs with active sensing devices to enable
the overall metasurface to estimate certain parameters of the impinging signal. However, this form of sensors-embedded RIS~\cite{RISsurvey2023} does not deploy its whole panel for sensing impinging signals, but only the part where its heterogeneous active sensing devices are located. In~\cite{hybrid_meta-atom}, a unit cell realizing simultaneous tunable reflections and signal absorption, via an embedded power splitter, was presented that constitutes the basic ingredient of Hybrid RISs (HRISs), which are also equipped with limited numbers of reception RF chains. These chains are attached via waveguides to disjoint groups of the latter dual-functional unit cells. Those metasurfaces have been lately deployed for boosting channel estimation in HRIS-assisted wireless communication systems~\cite{HRIS_CE} and enabling metasurface self-configurability~\cite{alexandropoulos2023hybrid}, localization~\cite{ghazalian2024joint}, as well as simultaneous communications and localization coverage in a given area of interest~\cite{HRIS_ISAC}. 

%
Despite the presented hybrid unit cells in~\cite{hybrid_meta-atom}, the design of dual-functional cells realizing fine-grained control over the phase, amplitude, and possibly the polarization of the reflected waves, while integrating sensing mechanisms to enable the estimation of critical environmental factors, including signal strength and channel state information, remains a challenging task especially for wireless application scenarios involving mobile users and rapidly fluctuating channel conditions. In this paper, we present a novel hybrid unit cell design comprising dual antenna elements, which share the same phase center, that supports both sensing and reflection functionalities, where the
one antenna is miniaturized via a dual dielectric material approach. By interleaving two arrays of hybrid unit cells of half-wavelength spacing, orthogonal polarization, and quarter-wavelength offset with a sub-wavelength reflecting array of split-ring unit cells integrated with a load-tuning matrix, we devise a novel HRIS architecture capable of sensing 
channel parameters simultaneously towards both communication ends wishing to profit from its reconfigurable reflections. Our full-wave-based performance evaluation results demonstrate the efficacy of both designed
hybrid unit cells and reflective split-ring ones comprising our noverl HRIS architecture.

\section{Proposed HRIS Design}\label{sec:Proposed_HRIS}
\begin{figure}[!t]
\centering
\includegraphics[width=1\columnwidth]{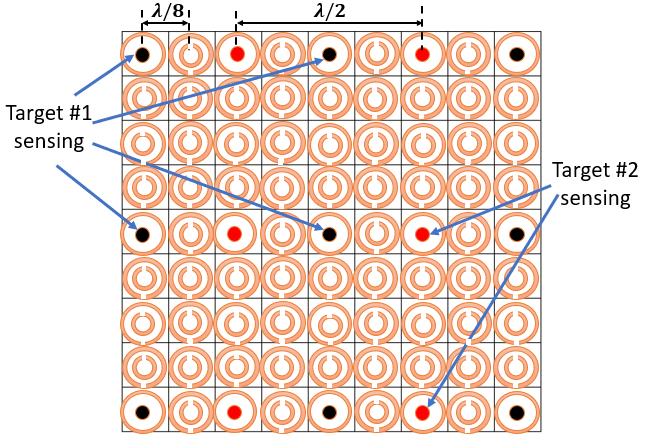}
\caption{The proposed HRIS architecture consisting of our hybrid unit cells for simultaneous sensing and reconfigurable reflections (Fig.~\ref{fig:hybridcell}) interleaved with split-ring unit cells offering additional reconfigurable reflections (Fig.~\ref{fig:Split}). The black and red hybrid unit cells indicate cell grouping into two separate signal feeders (e.g., reception RF chains) so as to enable dual sensing operation.}
\label{fig:array}
\end{figure}

The proposed HRIS architecture is depicted in Fig.~\ref{fig:array}. It comprises hybrid unit cells capable of simultaneous sensing and reconfigurable reflections (will be presented next in Section~\ref{sec:HUC}), which are interleaved with split-ring unit cells offering additional reconfigurable reflection (their design is described in Section~\ref{sec:RUC}). The black- and red-colored hybrid unit cells in the figure, which are all of size of one eighth of the free-space wavelength $\lambda$, indicate that these two groups of cells have been attached to separate signal feeders (e.g., reception RF chains) so as to enable dual sensing operation~\cite{alexandropoulos2020hardware,HRIS_CE}. In particular, the hybrid unit cells were interleaved within the HRIS architecture such that one sensing array senses parameters of the TX-HRIS channel (Target $\#1$), while the other senses the parameters of the RX-HRIS channel (Target~ $\#2$). The isolation between this sensing phased array pair can be successfully achieved through orthogonal polarization. As shown in the figure, the elements of the red/black colored groups are spaced almost $\lambda/2$ apart, while the solely reflective unit cells are spaced $\lambda/8$ apart. Both sensing arrays are interleaved with $\lambda/4$ spacing, ensuring that the maxima of each TX-HRIS sensing array element coincide with the minima of the RX-HRIS sensing array elements. Finally, the isolation between the hybrid unit cells and the reflective split-ring-based ones is achieved using different ground plane along with substrate, which reduces the wave propagation among the cells within the metasurface. Additional vias in circular pattern between a ring and a central patch can provide additional over-the-air leakage between the sensing and reflective unit cells.

\subsection{Hybrid Unit Cell}\label{sec:HUC}
\begin{figure}[!t]
\centering
\includegraphics[width=0.8\columnwidth]{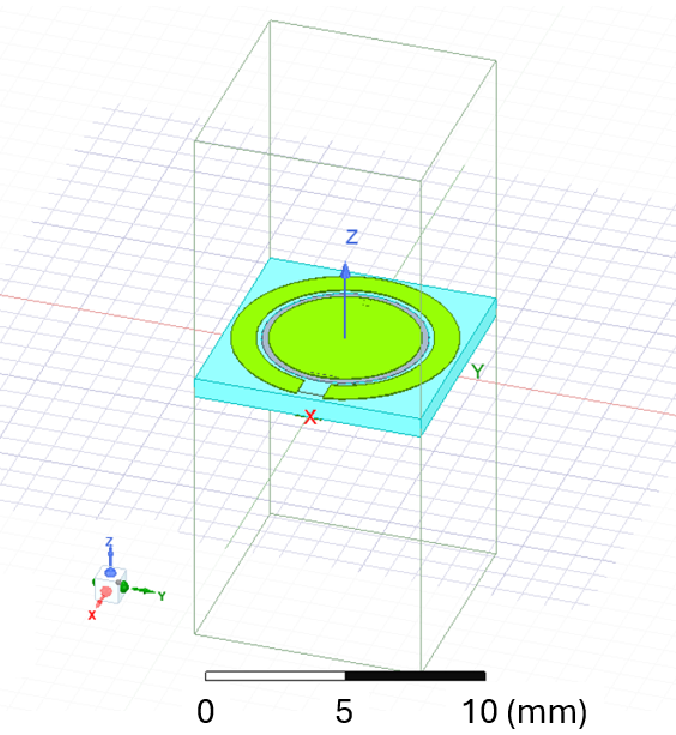}
\caption{The proposed hybrid unit cell of size $\lambda/8$ at $5.5$ GHz incorporating a ring antenna element with a central circular disc antenna of $\lambda_g/4$, capable of realizing simultaneous sensing and reconfigurable reflections.}
\label{fig:hybridcell}
\end{figure}
\begin{figure}[!t]
\centering
\includegraphics[width=.8\columnwidth]{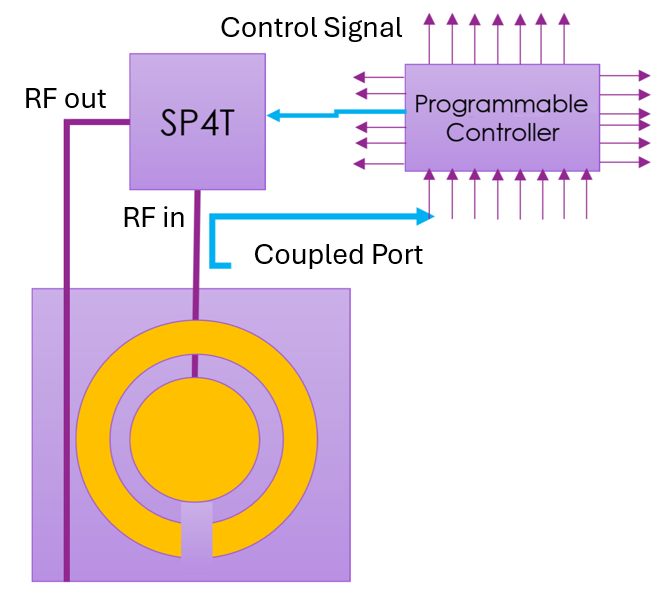}
\caption{The proposed hybrid unit cell with an SP4T switch and an integrated-circuit controller.}
\label{fig:Hybrid_Unit_Cell}
\end{figure}
The proposed hybrid unit cell design at $5.5$ GHz includes a ring element constructed from a low dielectric constant substrate together with a central circular disc antenna made from a high dielectric constant material, as illustrated in Fig.~\ref{fig:hybridcell}. The central circular disc antenna is responsible for sensing the phase and amplitude of incoming waves, while the ring antenna array reflects those waves in desired directions. The unit cell size is $7$ mm with an outer ring diameter of $6.4$~mm and an inner circular patch diameter of $3.8$ mm. The substrate thickness is $0.8$ mm and the copper thickness is $0.035$ mm. The dielectric material used for the outer ring antenna is Rogers ${\rm RO}4003$ with a dielectric constant of $\epsilon_r=3.5$, while the dielectric under the central circular disc is ${\rm RO}3010$ with a dielectric constant of $\epsilon_r=10.2$. We have deployed two dielectric materials to allow the central disc antenna to fit within the outer ring element. This implies that the antenna with an approximate size of  $\lambda_g/4=\lambda/(4\sqrt{\epsilon_r})$ will need to be less than the periodicity of the reflective unit cell that will be described in the sequel, which has been set to $\lambda/8$. It is, therefore, apparent that $\epsilon_r$ needs to be high enough to satisfy the above requirement, in particular, $\epsilon_r>4$. 

An additional shorting pin has been used in the central disc to further miniaturize the element size while maintaining the same resonant frequency~\cite{Kumar2003}. The introduction of this pin generates a parallel reactance, which, when combined with the admittance of the patch, effectively controls the resonant frequency. The impedance of the antenna is governed by the positioning of the shorting pin relative to the feed point. Specifically, moving the shorting pin away from the feed point increases the impedance, while moving it closer decreases it.

\begin{figure}[!t]
\centering
\includegraphics[width=0.8\columnwidth]{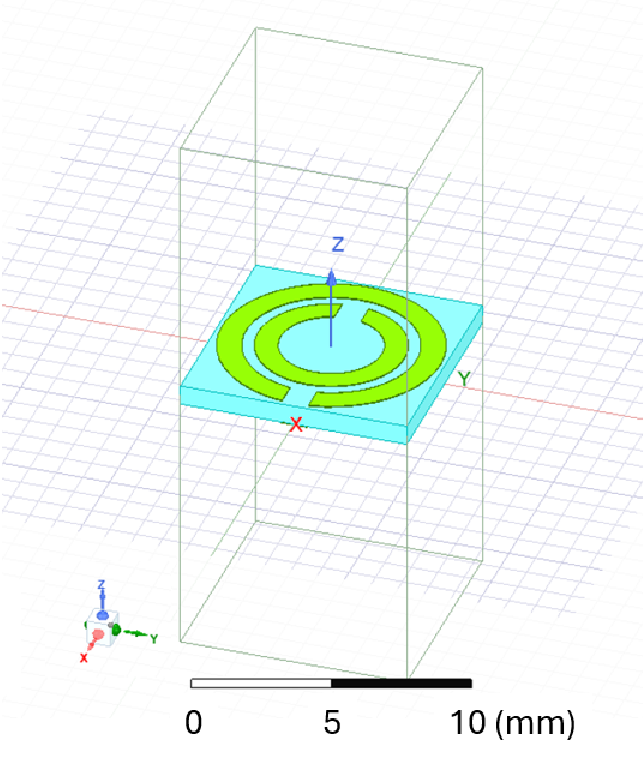}
\caption{The designed split-ring unit cell of size $\lambda/8$ at $5.5$ GHz capable of reconfigurable reflections.}
\label{fig:Split}
\end{figure}
The proposed hybrid unit cell is equipped with a solid-state SP4T switch~\cite{SP4T}. The common port of this switch is connected to the outer ring of each unit cell, while all four ports are connected to different loads. Figure~\ref{fig:Hybrid_Unit_Cell} illustrates how the SP4T switch and an integrated-circuit controller can be incorporated within the hybrid unit cell to enable sensing-aided tunability on a per hybrid-unit-cell basis. To this end, an incident wave impinges on the HRIS panel at an angle. The sensing elements of the metasurface (i.e., the respective parts of the hybrid unit cells) receive the signal and guide it to the controller, which is assigned to, e.g., compute the required phase shift. Through the control signal from the controller, the SP4T switch selects the appropriate delay to apply to the impinging signal, directing it towards the outer split-ring structure of the HRIS for reflection.

\subsection{Reflective Unit Cell}\label{sec:RUC}
In Fig.~\ref{fig:Split}, our designed unit cell that solely realizes tunable reflection of impinging waves is illustrated. This unit cell features a split-ring resonator on the ${\rm RO}4003$ substrate with a periodicity of $\lambda/8$. The split-ring resonator is anticipated to function similarly to the ring element of the hybrid unit cell, maintaining the same periodicity of $\lambda/8$. The reflecting unit cell is also equipped with an SP4T switch. 

The entire array, controlled by the SP4T switches, forms a load-tuning matrix. This matrix presents varying loads to the incident wave, effectively controlling the phase of the reflected wave. A lookup-table-based calibration is employed to identify the appropriate load configurations for directing the reflection. By sensing the direction of the target, the correct load matrix is applied to the reflecting array, ensuring that the reflected signal is steered towards the desired direction.

\section{Performance Evaluation Results}
\begin{figure}[!t]
\centering
\includegraphics[width=1\columnwidth]{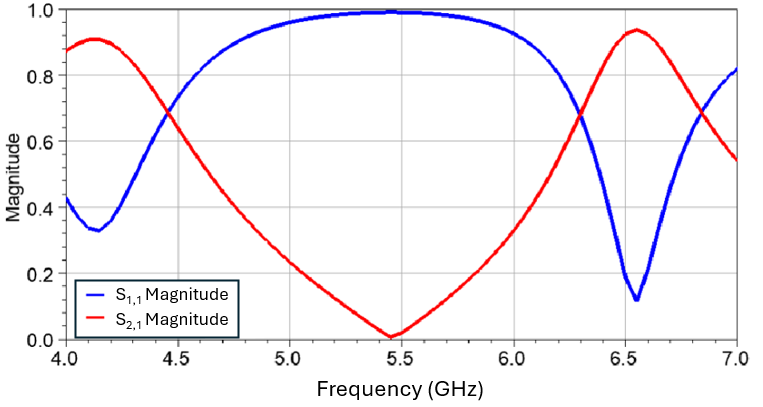}
\caption{The $S$-parameters of the designed hybrid unit cell using Floquet ports.}
\label{fig:SP}
\end{figure}
\begin{figure}[!t]
\centering
\includegraphics[width=1\columnwidth]{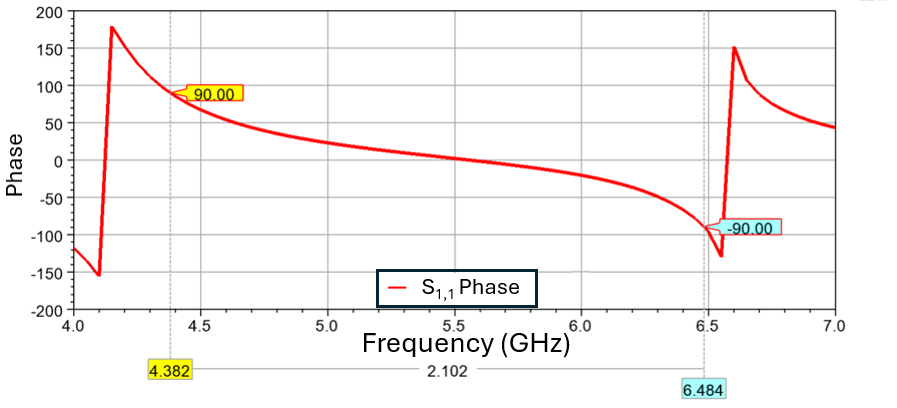}
\caption{The phase of the $S_{1,1}$ parameter of the designed hybrid unit cell.}
\label{fig:Phase}
\end{figure}
To accurately determine the permittivity and permeability of a metamaterial, it is crucial to extract the $S$-parameters directly from the realized structure. This approach provides a more precise characterization of the material's electromagnetic properties than analytical or theoretical approximations. However, the extraction of $S$-parameters necessitates the physical realization of the metamaterial, which involves an infinite repetition of its unit cells along the lattice vectors~\cite{Numan}. To simulate this periodicity, specific boundary conditions are typically applied during the design process. Two commonly used boundary conditions for this purpose are perfect electric and perfect magnetic boundary conditions. These conditions take advantage of the inherent symmetry of the metamaterial, effectively mimicking an infinite array of unit cells by creating a virtual periodic structure. By employing this method, the interaction between adjacent unit cells is accurately modeled, allowing for the precise extraction of $S$-parameters and, consequently, more reliable estimation of the material's properties~\cite{Andreone}.

The magnitude and phase of the $S$-parameters of the presented HRIS design in Section~\ref{sec:Proposed_HRIS} are illustrated in Figs.~\ref{fig:SP} and~\ref{fig:Phase}, respectively. The magnitude of $S_{1,1}$ represents the magnitude of the reflection coefficient at port $1$, quantifying how much of the incident wave on this port is reflected back when it interacts with the metasurface. A high value indicates significant reflection, while a low value suggests that the wave is either transmitted through the unit cell or absorbed. On the other hand, $S_{2,1}$ indicates how much of the wave entering through port $1$ is transmitted out through port $2$. A high magnitude suggests effective transmission, indicating that the metasurface is allowing the wave to pass through with minimal losses. It is noted that, when interpreting the results, impedance matching is key. If $S_{1,1}$ is low while $S_{2,1}$ is high, good impedance matching takes places. This means that the metasurface efficiently transmits energy rather than reflects it. In Fig.~\ref{fig:Phase}, it is demonstrated that the phase of the $S_{1,1}$ parameter is almost flat across the frequency $5.5$ GHz of interest, implying that the reflected wave will have no phase reversal.
\begin{figure}[!t]
\centering
\includegraphics[width=1\columnwidth]{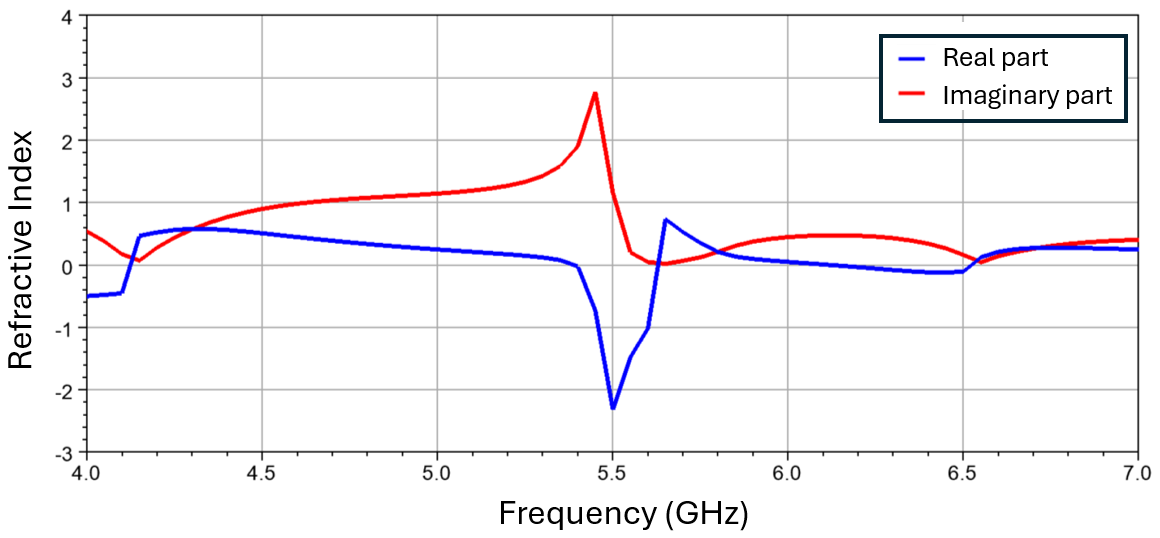}
\caption{Real and imaginary parts of the refractive index of the designed reflective split-ring unit cell.}
\label{fig:eta}
\end{figure}
\begin{figure}[!t]
\centering
\includegraphics[width=1\columnwidth]{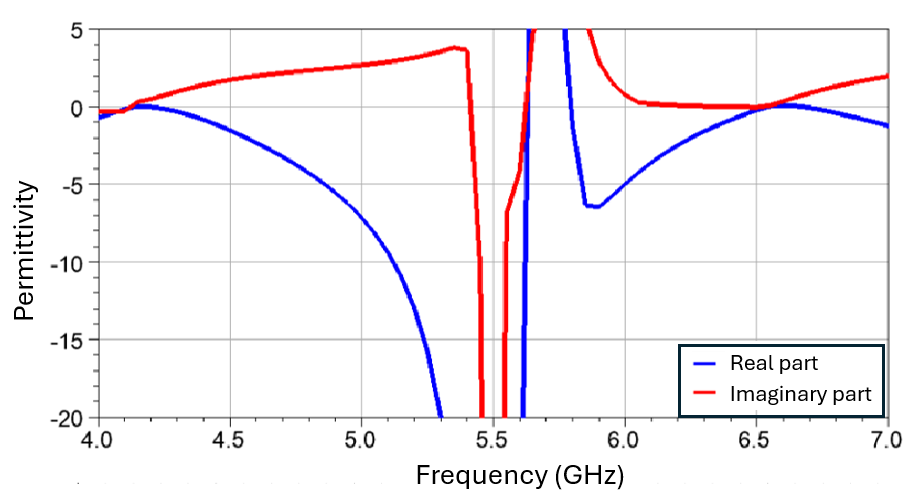}
\caption{Real and imaginary parts of the designed split-ring unit cell's permittivity.}
\label{fig:epsilon}
\end{figure}
\begin{figure}[!t]
\centering
\includegraphics[width=1\columnwidth]{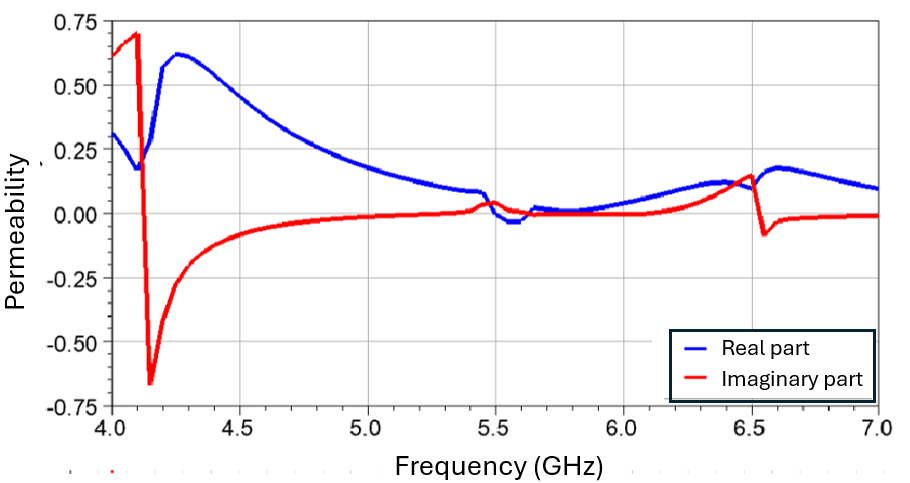}
\caption{Real and imaginary parts of the designed split-ring unit cell's permeability.}
\label{fig:Mu}
\end{figure}
Finally, the negative real refractive index of the designed reflective split-ring unit cell is illustrated in Fig.~\ref{fig:eta} confirming its metamaterial behavior~\cite{Numan}. As also depicted in Figs.~\ref{fig:epsilon} and~\ref{fig:Mu}, both permittivity and permeability exhibit negative values within the operating frequency range, closely aligning with the previous work~\cite{Smith} that identifies the material as a double-negative metamaterial. The negative real refractive index and the positive impedance observed around $5.5$ GHz further validate the metamaterial characteristics in this region.


\section{Conclusion}
In this paper, we presented a novel dual-functional HRIS design comprising dual substrate hybrid unit cells and reflective split-ring cells for simultaneous sensing and reconfigurable reflections. The design integrates two arrays of hybrid unit cells of half-wavelength spacing, orthogonal polarization, and quarter-wavelength offset with a sub-wavelength reflecting array of split-ring unit cells integrated with a load-tuning matrix, to enable sensing 
channel parameters simultaneously towards both communication ends wishing to profit from its reconfigurable reflections. Our full-wave simulations showcased the promising narrowband performance of the designed unit cells at $5.5$ GHz. However, further improvements are needed to scale our HRIS dual-functionality across all 6G frequency bands, paving the way for enhanced ISAC capabilities in future RIS-empowered wireless communication systems.

\section*{Acknowledgment}
This work has been supported by the SNS JU project TERRAMETA under the European Union’s Horizon Europe research and innovation programme under Grant Agreement No 101097101, including top-up funding by UKRI under the UK government's Horizon Europe funding guarantee.

\bibliographystyle{IEEEtran}
\bibliography{references}

\end{document}